\documentclass[useAMS,usenatbib]{mn2e}

\usepackage{times,graphicx,amssymb,natbib,color}

\usepackage{amsmath}
\title[Disc fragmentation rarely forms planets]{Disc fragmentation rarely forms planetary-mass objects} 
\author[Ken Rice, Eric Lopez, Duncan Forgan \& Beth Biller]{Ken Rice$^{1}$\thanks{E-mail: wkmr@roe.ac.uk}, Eric Lopez$^{1}$, Duncan Forgan$^{2}$  and Beth Biller$^{1}$\\
$^{1}$Scottish Universities Physics Alliance (SUPA), Institute for Astronomy, University of Edinburgh, Blackford Hill, Edinburgh, EH9 3HJ \\
$^{2}$Scottish Universities Physics Alliance (SUPA), School of Physics and Astronomy, University of St Andrews, North Haugh, St Andrews, KY169SS}

\begin{document}

\date{Accepted 0000}

\pagerange{\pageref{firstpage}--\pageref{lastpage}} \pubyear{0000}

\maketitle

\label{firstpage}
\begin{abstract}
It is now reasonably clear that disc fragmentation can only operate in the outer parts of protostellar discs ($r > 50$ au).  It is also expected that any object that forms via disc fragmentation will have an initial mass greater than that of Jupiter.  However, whether or not such a process actually operates, or can play a significant role in the formation of planetary-mass objects, is still unclear.  We do have a few examples of directly imaged objects that may have formed in this way, but we have yet to constrain how often disc fragmentation may actually form such objects.  What we want to consider here is whether or not we can constrain the likely population of planetary-mass objects formed via disc fragmentation by considering how a population of objects at large radii ($a > 50$) au - if they do exist - would evolve under perturbations from more distant stellar companions.  We find that there is a specific region of parameter space to which such objects would be scattered and show that the known exoplanets in that region have properties more consistent with that of the bulk exoplanet population, than with having been formed via disc fragmentation at large radii.  Along with the scarcity of directly-imaged objects at large radii, our results provide a similar, but independent, constraint on the frequency of objects formed via disc fragmentation.
\end{abstract}


\section{Introduction}
The most widely accepted mechanism for the formation of planets is the core accretion model \citep{pollack96}. In this model,
dust grains grow rapidly to form kilometre-sized planetesimals that then coagulate to form a rocky core \citep{safronov72} which,
if sufficiently massive, may then accrete a gaseous envelope to from a gas giant planet \citep{lissauer93, pollack96}. An alternative
suggestion \citep{kuiper51, boss98} is that gas giant planets may form via direct gravitational collapse in discs that are sufficiently
massive so as to sustain a gravitational instability \citep{toomre64}.  

The advantage of the latter mechanism is that it ensures that
gas giant planets can form prior to the dispersal of the gas disc, thought to typically occur within $\sim 5$ Myr \citep{haisch01}. Most
of the evidence, however, favours the standard core accretion mechanism. For example, gas giant planets are preferentially found
around metal-rich stars \citep{santos04,fischer05}, and there is an indication of a signature of the snowline in the exoplanet
distribution \citep{schlaufman09,rice13}. Both of these suggest that the amount of solid material in the disc, and the distribution of this material, 
influences planet formation, which would not be expected if disc fragmentation were a dominant formation mechanism.

Additionally, it is now fairly clear that disc fragmentation is physically implausible in the inner regions of protostellar discs \citep{rafikov05}. 
Fragmentation requires that the disc be both gravitationally unstable, and that it be able to cool rapidly \citep{gammie01,rice03}. 
The inner regions of protostellar discs are likely too optically thick to cool sufficiently fast for fragmentation to be possible 
\citep{clarke09,rice09}.  The outer regions (beyond $\sim 50$ au), however, may well have conditions suitable for fragmentation and it 
has, consequently, been suggested that this may explain some of the directly imaged exoplanets \citep{kratter10}, such as those in the HR8799 system \citep{marois08}. 

It has also been suggested \citep{nayakshin10a} that planets may form beyond $50$ au via disc fragmentation, and then spiral inwards, resulting in them orbiting much closer to their parent stars than where they formed.  Also, if both the grain sedimentation timescale and migration rate are fast, the planet may produce a core and lose some of its outer envelope through tidal interactions with the parent star \citep{nayakshin10b}.  As such, this process could turn what was originally a massive gaseous planet, into a lower mass Neptune-like planet, or even a terrestrial planet.  

Recent population synthesis calculations \citep{forgan13a}, however, suggest that this is very unlikely, and that it is much more likely that planets forming at large radii remain massive and remain with much larger radii than is typical for the known exoplanet population \citep{marcy08}. These initial population synthesis models also ignored subsequent mass accretion onto the planets and, hence, the masses are lower limits; we'd expect planets formed via disc fragmentation to have masses higher than these models suggest.  An initial study to also estimate how such a population would evolve through dynamical interactions suggests that some ($\sim 25$\%) would be ejected, with others scattered onto high-eccentricity orbits \citep{forgan15}.  However, this work did not consider the subsequent evolution of these high-eccentricity planets through tidal interactions with their parent stars.  

In this work we expand on \citet{forgan15} by considering how a population of planetary-mass bodies forming beyond $50$ au would evolve through dynamical interactions with an outer population of stellar companions, that drive Kozai-Lidov oscillations \citep{kozai62,lidov62}, and then through tidal interactions with the host star.  What we're aiming to do is to establish the orbital properties of planetary-mass objects that originated at large radii, and to determine if a population of such objects exists within the known exoplanet population.  In Section 2 we describe the models that we use.  In Section 3 we discuss our results, and we draw our conclusions in Section 4.

\section{Basic Model}
The goal is to study the evolution of a planetary system that is being perturbed by a third, stellar-mass body on an outer orbit.  Specifically, we aim to establish if we can identify a population, within the known exoplanet population, that could have originated with large initial semi-major axes ($a > 50$ au) and that has then been scattered onto orbits with semi-major axes inside $\sim 5$ au.  

To do this, we use the equations that describe the secular evolution of a star-planet-star system, first presented by \citet{eggleton01} and which have been extensively used to model the evolution of exoplanets that are perturbed by a third body \citep{wu03,fabrycky07}.   The equations include perturbing accelerations from a third body, general relativistic apsidal precession, perturbing accelerations from stellar and planetary distortions due to tides and rotation, and tidal interactions between the planet and its host star. We do, however, typically 
ignore the perturbing acceleration due to the planetary distortion as it only becomes significant when the planet is very close to its 
parent star, and reduces the timestep significantly. Here we use the form presented by \citet{barker09} and \citet{barker11} which are regular at $e = 0$, and we expand the contribution due to the third body to octupole order \citep{naoz11, naoz12}.  We also include a simple stellar wind model \citep{kawaler88} that allows the planetary host star to spin down as it evolves on the Main-Sequence.  Details of the model can be found in \citet{rice15}.  

\subsection{Initial conditions}
To understand how a population of planetary-mass bodies that form in the outer parts of protostellar discs via disc fragmentation, are perturbed by an outer stellar companion, we run a Monte Carlo-type simulation in which we randomly choose the outer planet to have a semi-major axis between $50$ and $80$ au, and to have an eccentricity that is initially small (drawn from the positive side of a Gaussian distribution with a half-width of $e = 0.025$).   We don't actually have a good sense of the initial properties for such a population of planets.  We could have selected the semi-major axis randomly in $\log a$, but that the range is quite small ($50 - 80$ au) probably means that this wouldn't make much difference, which is confirmed by a simple check of our results.  Also, the only simulations to consider this \citep{stamatellos09}  suggests that randomly in $a$ may be a more reasonable choice.  Similarly, other simulations \citep{hall15} suggest that the eccentricity of those objects that survive at large radii is typically small.  We also don't know the mass distribution for such a population of planets, but since we're interested in the possibility of planetary-mass companions forming at these radii, we assume the mass distribution is the same as that of the known population of exoplanets ($dN/dM_p \propto M_p^{-1.15}$) \citep{marcy08}.  The planet host star is assumed to have a mass of $M_* = 1$ M$_\odot$.   

That $30 - 40$\% of Solar-like stars have stellar, or sub-stellar, companions \citep{duquennoy91,raghavan10}, makes it likely that such companions will influence the evolution of the planetary system.  We, therefore, assume that there is a companion with a mass of $M_o = 0.5$ M$_\odot$ and with a semi-major axis, chosen randomly in log $a$, between $a_o  = 200$ and $a_o = 20000$ au \citep{duquennoy91}. Both 
radial velocity and transit exoplanet searches tend to remove binary systems from their catalogs. However, these tend to be systems where 
the separation is smaller than we've assumed here \citep{brown11a,valenti05} and so this should not bias our results. 

The companion's 
eccentricity is chosen randomly to be between $e_o = 0$ and $e_o = 1$, although we impose stability criteria \citep{mardling01, lithwick11, naoz13} that ensure that the triple system is long-term stable and that the quadrupole and octupole terms in the secular equations dominate (see \citealt{rice15} for details).  We could choose to vary the mass of this outer companion, but since we are in the test-particle regime \citep{lithwick11}, this should not influence the results.  We also fix the outer companion's orbit to be in the $xy$ plane and orient the inner orbit so that the mutual inclination, $i$, is distributed isotropically \citep{wu07}.  We also randomly orient the longitude of the planet's ascending node, and the argument of periastron.  

Since we're also interested in how the planet will evolve tidally with its host star (if scattered into an orbit that with a sufficiently small periastron) we assume that the star and planet have tidal quality factors of $Q'_s = 5 \times 10^6$ and $Q'_p = 5 \times 10^5$ \citep{goldreich66, yoder81, jackson08, baraffe10, brown11b}, and tidal love numbers of $k_s = 0.028$ and $k_p = 0.51$ \citep{petrovich15} respectively.  We also assume that any planet that reaches its Roche limit \citep{faber05}, given by,
\begin{equation}
 a = \frac{R_p}{0.462} \left( \frac{M_*}{M_p} \right)^{1/3}
 \label{eq:roche}
 \end{equation} 
is tidally destroyed.  

\section{Results}
As mentioned above, the goal is to establish if we can identify a population of exoplanets, from within the known population, that may have originated at large radii ($a > 50$ au) and that has then been scattered onto closer orbits via perturbations from an outer stellar companion that drives Kozai-Lidov oscillations. For our initial simulations, we randomly select - as described above - the properties of the inner and outer orbit and evolve the system, using a fourth-order Runge-Kutta integrator, for a randomly selected time of between 200 Myr and 10 Gyr.  We then repeat this to produce a sample of 5000 systems.   

\begin{figure}
\begin{center}
\includegraphics[scale = 0.45]{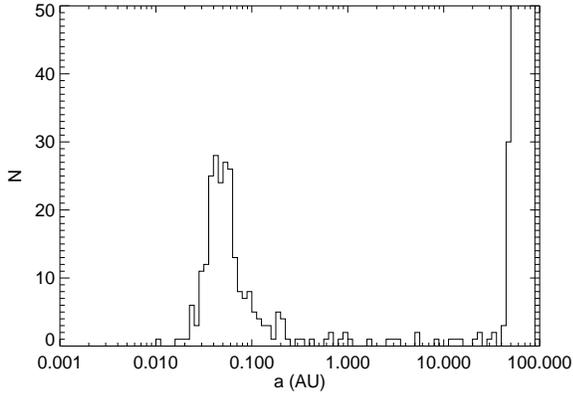}
\caption{The final semi-major axis distribution for a population of 5000 planets, initially located between $50$ and $80$ au and then perturbed by an isotropically distributed population of stellar companions with semi-major axes between $200$ and $20000$ au.  Most remain at large radii, but $\sim 5$\% are perturbed into orbits with $a < 1$ au.  A further $\sim 17$\% are perturbed into orbits with periastra inside their Roche limit and are, hence, tidally destroyed.}
\label{fig:aloghist1}
\end{center}
\end{figure}

Figure \ref{fig:aloghist1} shows the final semi-major axis distribution and illustrates how the perturbations from the outer stellar companion sculpts the initial semi-major axis distribution.  A large fraction - with a distribution that extends above the limits of the y-axis - remain with semi-major axes between $50$ and $80$ au.  Another population undergoes Kozai-Lidov oscillations which 
produces highly eccentric orbits that lead to these planets tidally interacting with their parent stars, and ending up on orbits 
with semi-major axes $a < 1$ au.  Another population (not shown in Figure \ref{fig:aloghist1}) reaches their Roche limit \citep{faber05} and is assumed to be tidally destroyed.
    
From an initial population of $5000$ systems, our simulations suggest that 864 ($\sim 17$\%) are tidally destroyed, $237$ ($\sim 5$\%) survive inside $1$ au, and the rest primarily remain out beyond $50$ au.  Binary companions to Solar-type stars are quite common, with 30-40\% having stellar, or sub-stellar, companions \citep{duquennoy91,raghavan10} with semi-major axes between $20$ and $20000$ au.  We've assumed that these companions have semi-major axes that are evenly distributed in log $a$ and have only considered companions with $a > 200$ au.  This would suggest that $20 - 30$ \% of Solar-type stars might satisfy these condition.  Given that these results suggest that $\sim 5$ \% of such systems could scatter an outer planet ($a$ initially between $50$ and $80$ au) into an orbit that gets tidally circularised - and survives - inside $a = 1$ au, would then suggest that 1 - 1.5 \% of all Solar-type stars could have such close-in planets if all such stars form planetary-mass companions beyond $50$ au via disc fragmentation. 

Admittedly, we've only presented results from a single set of parameters here.  However, we have run some tests with different $Q'_p$
and $Q'_s$ values and the results are broadly similar.  For very large values of $Q'_p$ ($Q'_p > 10^7$) the fraction surviving
inside $1$ AU can drop below $4$ \%, but that would still suggest that $\sim 1$ \% could have such close-in planets 
if companions that form at large radii via disc fragmentation are very common.

\begin{figure}
\begin{center}
\includegraphics[scale = 0.45]{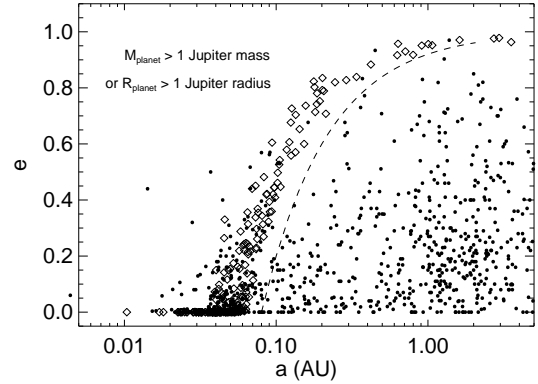}
\caption{Figure showing the final eccentricity, plotted against semi-major axis, for those planets that are scattered onto orbits - and survive - with $a < 5$ au.  The diamonds show our simulated planets, while the filled circles are the known exoplanets with masses above $1$ Jupiter mass, or radius above $1$ Jupiter radius, and with $a < 5$ au.  The planets scattered from large radii either become `hot' Jupiters, with $e \sim 0$, or proto-hot Jupiters.  This constrains the region of $e - a$ space where such planets could be found.}
\label{fig:aplaneouter1}
\end{center}
\end{figure}

Figure \ref{fig:aplaneouter1} shows the eccentricity, plotted against semi-major axis, for all those planets that have final semi-major axes inside $a = 5$ au (diamonds) and also shows (solid circles) all the known exoplanets with masses above $M_p = 1$ M$_{\rm Jup}$ or - for those without a mass estimate - with radii above $R_p = 1$ R$_{\rm Jup}$.  What it shows is that those planets scattered from beyond $50$ au will always end up either as `hot' Jupiters with $a < 0.1$ au and with small eccentricities, or as proto-hot Jupiters \citep{dawson12}.  Proto-Hot Jupiters are those gas-giant planets that are found on the high-eccentricity boundary, beyond which planets would typically be tidally destroyed, and that will ultimately tidally evolve to become `hot' Jupiters.   This is quite useful in that it constrains the region of $e - a$ space where we would find such planets, if they do indeed exist. 

Again, we've only presented results from a single set of parameters. We've used a relatively low $Q'_p$ value and we find
that higher $Q'_p$ values tend to result in proto-hot Jupiters having larger eccentricities, for a given final semi-major axis. In fact, for sufficiently large $Q'_p$ values ($Q'_p > 10^7$) the proto-hot
Jupiters end up in a region of $e - a$ space where very few exoplanets are found.  It does seem clear, therefore, that if planetary-mass 
bodies are scattered from very large initial radii, they would likely be found either as `hot' Jupiters or as proto-hot Jupiters, near
the boundary beyond which they would probably be tidally destroyed.

Figure \ref{fig:aplaneouter1} introduces the first possible issue with this mechanism playing a significant role in producing close-in exoplanets.  As mentioned above, if planet formation via disc fragmentation is common (i.e., most stars have gas giants orbiting beyond $50$ au) then $1 - 1.5$ \% of stars could have close-in exoplanets that were scattered in from these large initial radii.  However, Figure \ref{fig:aplaneouter1} shows that these would have to be either `hot' Jupiters, or be proto-hot Jupiters undergoing high-eccentricity tidal migration.  Currently, only about $1$ \% of Sun-like stars have such planets \citep{fressin13, petigura13, dawson15} and, so, if planet formation via disc fragmentation is common, this scattering scenario could explain almost all the known `hot' and proto-hot Jupiters.

However, one can also largely explain the `hot' Jupiters and proto-hot Jupiters through a combination of disc migration and scattering (both from other planets and from stellar and sub-stellar companions) of planets that initially form at modest semi-major axes (inside $\sim 10$ au) \citep{nagasawa08}.  In fact, we reran our simulations with the only change being that the planets were randomly distributed between $5$ and $10$ au, rather than between $50$ and $80$ au.  In this case, just over $5$ \% were scattered onto close-in orbits, either `hot' Jupiters or proto-hot Jupiters.  This is possibly a slight under-estimate as we've restricted the companions to be beyond $200$ au.  However, if $20$ \% of Sun-like stars have gas giant companions \citep{marcy08}, and $20 - 30$ \% of such systems have stellar companions, then this would suggest that $0.2$ - $0.3$ \% of such systems will end up as `hot' Jupiters or proto-hot Jupiters via this mechanism alone.  Moreover, when \citet{dawson15} analyzed the distribution of transit durations, and therefore eccentricities, among `hot' Jupiters found by {\em Kepler}, they found that eccentric proto-hot Jupiters are extremely rare compared to `hot' Jupiters, indicating that the majority of the observed `hot' Jupiters likely arrive by disc migration and not by high eccentricity scattering from the snow-line or beyond. Therefore, that closer stellar companions, other planetary companions, and disc migration could also play a role in forming such planets, suggests that such mechanisms are sufficient to explain the observed population, and that scattering planets from beyond $50$ au is unlikely to play a significant role.

\subsection{The properties of close-in exoplanets}
\begin{figure}
\begin{center}
\includegraphics[scale = 0.45]{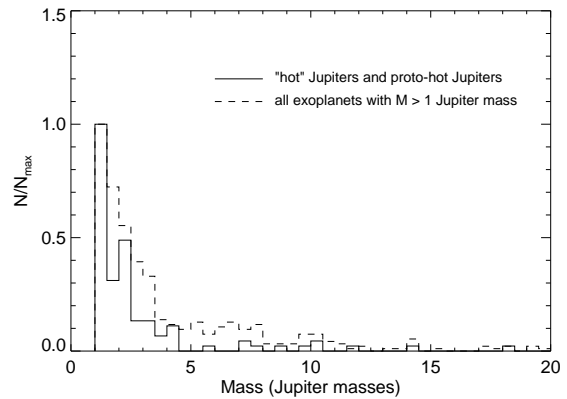}
\caption{The solid-line shows the mass distribution of all known exoplanets with masses above $1$ Jupiter that exist in the region to which outer planets could be scattered (found to left of the dashed-line in Figure \ref{fig:aplaneouter1}). The dashed-line shows the mass distribution of all known exoplanets with masses above $1$ Jupiter mass.  It's clear that the two distributions are very similar and is consistent with `hot' Jupiters and proto-hot Jupiters coming primarily from the bulk exoplanet population.}
\label{fig:exomasshist}
\end{center}
\end{figure}

One way to further investigate this is to consider the properties of the known exoplanets that lie within the region to which these outer planets could be scattered.  Figure \ref{fig:aplaneouter1} suggests that planets scattered from beyond $50$ au can only become either `hot' Jupiters, or proto-hot Jupiters.  Consequently, we assume that only those known exoplanets lying to the left of the dashed line in Figure \ref{fig:aplaneouter1} could have been scattered from initial radii beyond $50$ au. As mentioned above, the exact location in $e - a$ space
to which proto-hot Jupiters could be scattered does depend somewhat on the tidal dissipation parameters. However, using a higher 
$Q'_p$ value would likely move the boundary to slightly higher $e$ values, and so our choice of boundary (dashed line in Figure \ref{fig:exomasshist}) should at least capture most objects that could have been scattered from a large initial semi-major axis. 

Figure \ref{fig:exomasshist} shows the mass distribution for the known exoplanets with masses above $1$ Jupiter mass that exist to the left of the dashed-line in Figure \ref{fig:aplaneouter1} (solid line) and also shows the mass distribution of all known exoplanets with masses above $1$ Jupiter mass.  Both distributions are normalised with respect to their maximum value.

The two distributions are very similar and is therefore consistent with most `hot' Jupiters and proto-hot Jupiters coming from the main exoplanet population.  Additionally, most of the planets have masses below $5$ Jupiter masses, and more than $65$\% have masses below 3 Jupiter masses.  It has been suggested \citep{forgan11} that the initial mass of planets that form via disc fragmentation will typically exceed $3$ Jupiter masses.  Even after tidal downsizing, the vast majority of objects still have masses above 5 Jupiter masses \citep{forgan13a} and, if there is any external irradiation, the initial mass increases \citep{forgan13b}.  However, a large fraction of the observed `hot' Jupiters and proto-hot Jupiters have masses below $3$ Jupiter masses and, therefore, have masses below that expected for planets that formed via gravitational instability.  

As discussed earlier, our simulations assumed that the objects forming at large initial radii had the same mass distribution as the 
bulk exoplanet population.  Since it seems likely that objects forming via disc fragmentation would tend to have initial masses of a few
Jupiter masses, this may not be an appropriate mass distribution.  However, even if we repeat our simulations with the mass fixed
at $15$ M$_{\rm Jup}$, the fraction that survive inside $1$ au drops from $\sim 5$ \% to $\sim 2.5$ \%, and the region of $e - a$ space
to which they are scattered remains unchanged. Therefore, even if our mass distribution is not quite representative of what would
be expected for objects forming at large radii via disc fragmentation, this wouldn't change our results significantly. 

It is also generally recognised that the host stars of gas giants tend to be metal-rich compared to a typical sample of similar stars \citep{santos04, fischer05}.  This is thought to indicate that gas giant planets are more likely to form in discs that are enhanced in solids and is regarded as consistent with the standard core accretion scenario \citep{pollack96}.  Disc fragmentation would not seem to require this enhancement and, if anything, it may even be more effective in systems that are metal poor \citep{meru10, clarke09} since such discs can cool more efficiently than discs that are enhanced in solids.   It has been suggested, however, that there may be a similar metallicity dependence for planets that form via gravitational instability, undergo pebble accretion, and then migrate rapidly to the inner disc \citep{nayakshin15}.  This, however, does not necessarily apply to those that still have large semi-major axes once the disc has dissipated, and it is this population that is relevant to what we're presenting here. 

\begin{figure}
\begin{center}
\includegraphics[scale = 0.45]{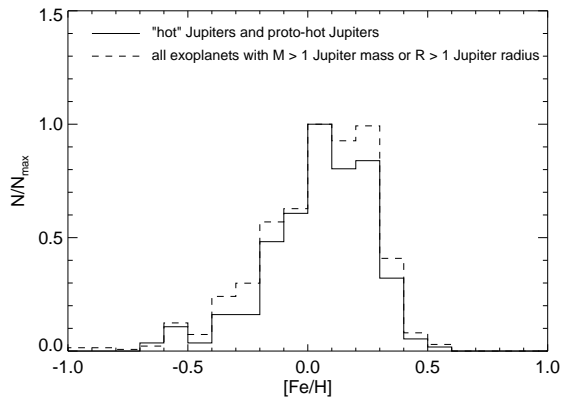}
\caption{The solid-line shows the metallicity distribution for all known exoplanets - with masses above $1$ Jupiter mass, or radii
above $1$ Jupiter radius - that exist in the region to which outer planets could be scattered (to the left of the dashed-line in Figure \ref{fig:aplaneouter1}.  The dashed-line is the metallicity distribution for all known exoplanets with masses, or radii, greater than that of
Jupiter.  That both populations are preferentially found around metal-rich stars is consistent with `hot' Jupiters and proto-hot Jupiters coming primarily from the bulk exoplanet population.}
\label{fig:exoFeHhist}
\end{center}
\end{figure}

Figure \ref{fig:exoFeHhist} shows the metallicity distribution for those known exoplanets in the region to which outer planets can be scattered (solid line) compared to all known exoplanets (dashed line).  In both cases, we restrict the mass to be above $M = 1$M$_{\rm Jup}$ or - for those without a mass estimate - the radius to be above $R = 1$ R$_{\rm Jup}$. It's clear that the distributions are very similar and that - like the bulk exoplanet population - `hot' Jupiters and proto-hot Jupiters tend to be found around metal-rich stars.  Again, this is consistent with such planets being scattered, or migrated, from the bulk exoplanet population, rather than from radii beyond $50$ au.  

To further quantify if the `hot' and proto-hot Jupiters could have originated from the bulk exoplanet population, we carried out Kolmogorov-Smirnoff tests on the metallicity and mass distributions.  A Kolmogorov-Smirnoff test of the metallicity distribution returns a
probability of $P_{KS} = 0.78$, indicating that we can't
rule out that `hot' and proto-hot Jupiters are drawn from the same distribution as the bulk exoplanet population 
(dashed-line in Figure \ref{fig:exoFeHhist}).  Similarly, the KS test of the mass distributions (Figure \ref{fig:exomasshist}) returns
a probability of $P_{KS} = 0.22$, again indicating that we can't rule out that the two populations are drawn from the same
distribution. 

Figure \ref{fig:exomasshist} therefore suggests that most `hot' Jupiters and proto-hot Jupiters have lower masses than would
be expected for planets forming via disc fragmentation, and the KS test indicates that we can't rule out that they're drawn from the same
distribution as the bulk exoplanet population. This would appear inconsistent with a significant fraction originating 
from an outer population that formed via disc fragmentation.  That we also can't rule out that the metallicity distribution
is drawn from the same distribution as the bulk exoplanet poopulation would also suggest that `hot' and proto-hot Jupiters are
more likely to have come from the main exoplanet population, than from a population of planetary-mass bodies that formed via disc
fragmentation at large orbital radii.

\subsection{KELT1-B - The closest brown dwarf companion}
Possibly one of the more interesting objects in the context of what we're considering here is KELT-1B \citep{siverd12}.  It is a brown dwarf, rather than a planetary-mass, companion to a $1.335$ Solar-mass star on an orbit with a semi-major axis of $a = 0.024$ au, and an eccentricity of $e = 0.001$.  It has a mass of $27.4$ Jupiter masses and so is more consistent with what might be expected for objects that form via disc fragmentation at large radii \citep{stamatellos09, forgan13a, forgan13b}.  We didn't directly address this system, as it is beyond what we can consider here, but we did repeat our simulations with the only changes being that we simulated 2500 systems (rather than 5000) and assumed that the outer non-stellar object had a mass of $30$ Jupiter masses.  

\begin{figure}
\begin{center}
\includegraphics[scale = 0.45]{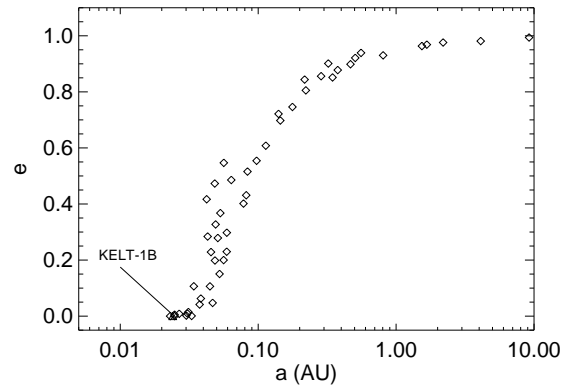}
\caption{Figure showing the final eccentricity, plotted against semi-major axis, for objects with a mass of $M = 30$ Jupiter masses, that started with a semi-major axis between $a = 50$ and $a = 80$ au, and that were scattered by an even more distant stellar companion.  As with the simulations of planetary-mass objects, the only objects that survive inside $a = 10$ au are those that end up inside $a = 0.1$ au and have a small eccentricity, or those that are on the high-eccentricity boundary, beyond which they would be tidally destroyed.  The location of KELT-1B is indicated at $a = 0.024$ au and $e = 0.001$, showing that scattering from a large initial radius could potentially form such a system.}
\label{fig:aplaneouter30MJup}
\end{center}
\end{figure}

As with our earlier simulations, a large fraction of the objects are scattered onto orbits that pass close to the parent star.  In the simulations with an outer object with $M = 30$ M$_{\rm Jup}$, 558 ($22$ \%) are tidally destroyed and 45 ($1.8$ \%) survive on orbits with $a < 1$ au.  This means that a smaller percentage survive with small semi-major axis, than for the case where the outer object was assumed to be of planetary-mass, but a similar fraction are scattered. Again, as shown in Figure \ref{fig:aplaneouter30MJup}, they're all either in the same location as `hot' Jupiters or proto-hot Jupiters, beyond which they would be tidally destroyed.  The position of KELT-1B is indicated at $a = 0.024$ au and $e = 0.001$.  As is clear, it is possible for a system such as KELT-1B to form via scattering from a large initial radius.   

As already mentioned, about $1$ \% of Sun-like stars have a `hot' Jupiter or a proto-hot Jupiter \citep{fressin13,petigura13,dawson15}.   If we consider the known exoplanets with mass estimates, there are 127  with masses above $1$ Jupiter mass and that are either `hot' Jupiters or proto-hot Jupiters (i.e., lie to the left of the dashed line in Figure \ref{fig:aplaneouter1}).  If we also consider those without mass estimates, but with radii above 1 Jupiter radius, it increases to 282.  KELT-1B is one of the few brown dwarf companions that have orbital properties similar to that of a typical `hot' Jupiters.  The only other known one is Corot-3b \citep{triaud09}, a $21$ Jupiter mass companion to a $1.36$ Solar-mass star, with an orbital semi-major axis of $a = 0.057$ au.   As mentioned above, a $30$ Jupiter mass object, scattered from a large initial radius, is a few times less likely to survive inside $a = 1$ au, than a planetary-mass body.  Consequently, if KELT-1B is indicative of a population of bodies at large orbital radius, a few of the known `hot' and proto-hot Jupiters may have originated via scattering from such a population.  In other words, maybe $\sim 1$\% of the known `hot' and proto-hot Jupiters originated from beyond $\sim 50$ au. However, as mentioned earlier, all the known `hot' and proto-hot Jupiters could have such origins if just over half of Solar-like stars have a population of outer planetary companions.  Therefore, if only $\sim 1$\% of known `hot' and proto-hot Jupiters have such an origin, less than $\sim 1$\% of Sun-like stars can have planetary, or brown dwarf, mass companions at large radii.  

\subsection{Properties of wide exoplanets probed by direct imaging}
As we've shown above, planetary-mass bodies scattered from large initial radii ($a > 50$ au) can only be found as `hot' Jupiters or proto-hot Jupiters.  We've also shown that the properties of the known exoplanets in those regions are very similar to that of the bulk exoplanet population, suggesting that very few - if any - were scattered from beyond $50$ au.  A simplistic analysis based on the existence of a KELT-1B, a brown dwarf with orbital properties similar to that of `hot' Jupiters, does suggest, however, that maybe a small number of the known `hot' and proto-hot Jupiters could have originated beyond $\sim 50$ au.  This would indicate that maybe $\sim 1$\% of Solar-like stars have distant planetary, or brown dwarf, companions.  

Most planetary detection methods (e.g. radial velocity and transit, for instance) are, however, more sensitive to close-in planets ($<$5 AU) than to those that may be
further out. In contrast, direct imaging is more sensitive to planets at wider separations, and of all planetary detection methods uniquely
probes planets at large radii ($>$50 au).  To date, $\sim$20 wide exoplanet or very-low-mass objects ($<$25 M$_{Jup}$) have been discovered
(\citealt{marois08, marois10, lagrange09, lagrange10, kuzuhara13}, among others).  In the last decade, numerous 8m telescope
+ coronagraph surveys have searched for planetary-mass bodies around young stars (e.g. \citealt{lowrance05, biller07,
lafreniere07, chauvin10, heinze10, vigan12, biller13, rameau13, wahhaj13, chauvin15, brandt14, bowler15}), in order to detect new companions and also to determine the frequency
of such wide companions.  Although individual surveys tend to only cover 80-100 stars, these surveys have - overall - now observed a sample of $\sim$500 stars. 

They have unearthed few new planets and most have yielded null planet detections, leading to the unambiguous
result that planetary-mass bodies are rare at large radii ($>$50 AU).  For instance, Biller et al. (2013) find, from a Bayesian analysis of a null planet
detection from 78 young moving group stars observed with the Gemini NICI planet-finder, that the frequency of 1-20 M$_{\rm Jup}$ companions
at semi-major axes from 10-150 AU is $<$18\% at a 95.4\% confidence level using DUSTY models (Chabrier et al. 2000) and is $<$6\% at
a 95.4\% confidence level using COND models (Baraffe et al. 2003).  Brandt et al. 2014b combine imaging data from multiple surveys to
build a 250 star survey with 5 detections of brown dwarf companions.  Modelling this population with a single power-law distribution, they
find that 1-3$\%$ of stars (68$\%$ confidence) host 5-70 M$_{\rm Jup}$ companions between 10-100 AU.  They argue that this suggests
that most wide companions formed through disc fragmentation rather than through core accretion.  However, whichever formation mechanism
is responsible for their formation, wide giant-exoplanet companions are quite rare.

Further analysis, however, suggests that the few known wide companions are indeed more likely to have been formed by disc fragmentation, than via core accretion.  Biller et al. 2013 marginalize over a
wide range of potential planetary distributions and Brandt et al. 2014b explicitly fit power-laws unconnected to radial velocity studies. 
However,  most statistical treatments of directly imaged planetary companion distributions adopt fixed power-law distributions in semi-major
axis and planet mass drawn from radial velocity planet studies (e.g. Cumming et al. 2008) and extend these proscriptions to separations
appropriate for directly imaged planets (e.g. $>$10 AU).  As the population of planets probed by radial velocity is likely formed by
core-accretion \citep{fischer05}, these simple empirical power-laws have been interpreted (even when extended to wide separations)
as the expected population as produced by core-accretion.  Since the power-laws used in radial velocity studies extend to arbitrary
semi-major axes, these studies add a new ``cutoff" parameter, truncating the power-law distribution at a specific semi-major axis
(assumed to be similar to the size of the primordial disc).  These studies have found that the semi-major axis cutoff for core-accretion
planets must be $<$75 AU (e.g. Nielsen and Close 2010) and likely less than 30-40 AU.

Although this analysis suggests that planetary-mass bodies at large radii probably formed via disc fragmentation, rather than via core accretion, the direct imaging surveys are also consistent with the results of the simulations presented here: planetary-mass bodies at large radii are rare and, therefore, disc fragmentation rarely forms such objects.   It is possible that a few percent of stars host such companions, but that is consistent with the suggestion in Section 3.2, that at most a few of the known `hot' Jupiters and proto-hot Jupiters could have been scattered from a large initial radius.

\section{Conclusions}
In this study we have aimed to investigate if we can identify a population, from within the known exoplanet population, that may have originated from large initial radii ($a > 50$ au).  Given that a reasonable fraction ($\sim 30 - 40$  \%) of Solar-like stars have stellar companions, a population of planetary-mass companions beyond $50$ au - if it does exist - should be susceptible to scattering via perturbations from these even more distant stellar companions.  The analysis here suggests that if any such planets are scattered to closer orbits they should be found as `hot' Jupiters ($a < 0.1$ au and $e \sim 0$) or proto-hot Jupiters; gas-giant planets on the high-eccentricity boundary, beyond which they would typically be tidally destroyed.  However, when we consider the known exoplanets that lie in that region of parameter space, they have properties (mass and metallicity distributions) that are consistent with that of the bulk exoplanet population.  Similarly, the typical mass of the known exoplanets in these regions is less than that expected for planets that form via disc fragmentation at large radii.  Additionally, direct imaging searches for planetary-mass bodies around Solar-like stars also suggest that such objects are rare.  Therefore, we conclude that there is probably not a substantial population of planetary-mass bodies at large radii around Solar-like stars and, consequently, that disc fragmentation rarely forms such objects.

Of course, there are a number of caveats to the above.  We only presented results from a single set of parameters in our simulations, but most were chosen to be conservative.  For example, the stellar wind parameter was chosen to spin the star down slightly more than may be reasonable.  A weaker wind would remove less angular momentum, slow the tidal evolution of the planet, and allow a larger fraction to survive.  The tidal quality factors were also chosen to be on the low side of the range.  Additional tests suggest that larger $Q'_p$ values wouldn't 
change the fraction surviving inside $1$ au substantially and would tend to move the proto-hot Jupiters to even larger eccentricities
than considered here. Therefore, if planetary-mass objects at large radii were common, they could explain a large fraction of the known `hot' Jupiters and proto-hot Jupiters.   That such known exoplanets appear more consistent with an origin from within the main exoplanet population suggests that there is not a substantial population of planetary-mass bodies at large radii.  

As suggested above, a possible consequence of this analysis is that disc fragmentation rarely forms planetary-mass bodies.  One possibility is that any object formed via this mechanism would quickly grow in mass to become a brown dwarf ($M > 15$ M$_{\rm Jup}$) as suggested by disc fragmentation simulations \citep{stamatellos09}.  An example might be KELT-1B, which has a mass of $M = 27.4$ M$_{\rm Jup}$ and an orbital radius of $a = 0.024$ au.  Our analysis suggests that it may indeed have been scattered from a large initial orbital radius. Additionally,  
KELT1-B is one of only a small number of known brown-dwarf mass objects that lie in the region of parameter space to which we'd expect outer objects to be scattered.  Therefore, even if KELT-1B is an example of an object scattered from beyond $50$ au, it would still be consistent with such objects being rare and, hence, that disc fragmentation rarely operates.  

Studies of self-gravitating disc evolution \citep{whitworth06, rice09, clarke09, forgan11} do, however, suggest that any disc with an initial outer radius that extends beyond $\sim 50$ au should be susceptible to fragmentation and, consequently, the formation of planetary or brown dwarf-mass objects.  That disc fragmentation seems unlikely, may suggest that discs with such large initial radii are rare, or that external irradiation is able to stabilise such discs against fragmentation \citep{rice11}.  There are indeed observations \citep{maury10} suggesting that discs around in very young Class 0 protostellar systems do tend to be compact, with outer radii inside $50$ au.  If so, then by the time such discs have expanded to larger radii, there may be insufficient mass for them to be susceptible to fragmentation \citep{rice10}.  There is, however, some evidence for extended discs in very-young protostellar systems \citep{tobin12}, which may be gravitationally unstable \citep{forgan13c}.  If such discs do exist, then the possibility that disc fragmentation rarely operates, may indicate that something, such as magnetic fields \citep{commercon11}, is inhibiting fragmentation in such discs.  

Additionally, fragment destruction is a common outcome for all disc fragmentation studies, especially as self gravitating disc migration appears to be relatively rapid, and induces tidal disruption before the fragment is fully bound \citep{baruteau11, zhu12, forgan13a}.  Dynamical
evolution while still in the nascent stellar cluster, could also remove objects that survive at large radii \citep{forgan15}.  This, however,
would need to be very efficient if it were to reduce an initially large population of outer planetary-mass bodies, to one more consistent 
with the results presented here.  It is possible, therefore, that the results here have implications for planet formation, suggesting that most known exoplanets formed via core accretion, and for star formation itself, suggesting - for example - that discs around very young stars are typically compact, with initial outer radii inside the radius where disc fragmentation becomes viable,
or that magnetic fields and external irradiation can stabilise radially extended discs against fragmentation.  

\section*{Acknowledgements}
The author acknowledges very useful discussions with Adrian Barker, Gordon Ogilvie, Phil Armitage and Douglas Heggie. KR gratefully acknowledges support from STFC grant ST/M001229/1.  The research leading to these results
also received funding from the European Union Seventh Framework Programme (FP7/2007-2013) under grant agreement number
313014 (ETAEARTH).  DF grateful acknowledges support from the ECOGAL ERC advanced grant.

\label{lastpage}

\end{document}